\documentstyle[12pt,rotate]{article}
\newcommand{\frat}[2]{\frac{\textstyle #1}{\textstyle #2}}

\newcommand{\vf}[1]{\mbox{\boldmath $#1$}}
\newcommand{\grpicture}[1]

\begin{document}
\begin{center}
{\large\bf Simple Mechanism of Softening Structure Functions at Low Transverse
Momentum Region}\\[0.5cm]
{S.V.Molodtsov$^1$, A.M.Snigirev$^2$, G.M.Zinovjev$^{3,4}$}\\[0.5cm]
{\small\it $^1$ State Research Center Institute of Theoretical and Experimental Physics,\\
117259 Moscow, RUSSIA\\
$^2$ Nuclear Physics Institute, Moscow State University,\\
119899 Moscow, RUSSIA\\
$^3$Fakultat fur Physik, Universitat Bielefeld,\\
33615 Bielefeld, GERMANY\\
$^4$Bogolyubov Institute for Theoretical Physics,\\
National Academy of Sciences of Ukraine,\\
252143 Kiev, UKRAINE}
\end{center}
\thispagestyle{empty}
\abstract

The relevance of the dipole configurations of quarks in forming 
nucleus structure functions is discussed. It is shown that a radiation
generated by dipole configurations while moving relativistically along
their axises is described by distributions which are finite and infrared
stable in low transverse momentum region. It is argued that there is an 
exponential transition  to the perturbative regime of large transverse
momenta and its power is defined by the distance between the dipole
charges in its rest frame $m^{-1}_\pi$. 
\newpage

\section{Introduction}

Theory of dense relativistic gluon (parton) systems is recently recognized
to be crucially important solving many (even long time pending) problems 
of high energy hadron physics \cite{1}. But on eve of new round
of the experiments with ultrarelativistic nuclear beams (at RHIC and LHC)
this subject becomes a real challenge because of an extreme necessity to
answer most reliably the question about the initial conditions and early
evolution of the quark-gluon plasma (if it is formed) as emphasized in
\cite{2}. Apparently clear, an answer depends essentially on the behaviour
of the nuclear structure functions which are calculable perturbatively in
QCD \cite{3}. However, extrapolating these results even to the region of
moderate parton transverse momenta takes very special efforts \cite{4} and
leads to the singular distributions for the small momentum values.

Moreover, a rather provocative idea in this context launched in \cite{5}
suggests for the largest nuclei at very high energy the initial parton
density could become so high that the intrinsic transverse parton momenta
reach the magnitudes where an overlap with the region mastered by
perturbative QCD already occurs. It treats the structure function of
nucleus as an aggregate of quantum fluctuations on the ground of classical 
gluon fields generated by ultrarelativistic (anti-)quarks and argues that 
respective classical beam bremsstrahlung could modify many results which
are discussed as possible quark-gluon plasma signatures. In the course of
such calculations \cite{5}, \cite{6} light cone QCD is handled in the
light cone gauge \cite{7} because of taking the limit of $v \to 1$ becomes 
quite nontrivial problem \cite{8} and the corresponding potentials turn
out pretty singular and having no the direct physical meaning.

In the present paper dealing with classical electrodynamics in order to
make the results somewhat easily understandable we show that the field
strengths of particular dipole configurations regularized by averaging
$E_{reg}$, $ H_{reg}$ justify taking the immediate limit of $v \to 1$ for
physical observables and provide reasonable asymptotic behaviour for the 
structure functions. We believe it could be indicative to appraise
nonperturbative region role contributing to the behaviour of nuclear
structure functions in the McLerran-Venugopalan approach.

The paper is simply structured. In Sect.\ 1 we mainly remind the basic
formulae for the potentials and field strengths of ultrarelativistic
particles in classical electrodynamics. The radiation of a fast-moving
classical charge in the fields considered is calculated in Sect.\ 2. We
conclude discussing the relevance of these results to describe the nuclear
structure functions in the final Sect.\ 3.

\section{Field of ultrarelativistic particles in electrodynamics}

If one considers the field of a classical charge $e$ freely moving with
the constant velocity $v$ the following expression for the potential is
valid demonstrating obviously the cylindrical symmetry of a solution in
the Minkowski space ($c=1$) 

\begin{equation}
\label{eq1}
A_\mu(x)=\frat{e~u_\mu}{R^{'}},
\end{equation}
where $u_{\mu}=(u_0,~{\vf u}),~{\vf u}=vu_0{\vf n},~
u_0=\frat{1}{(1-v^2)^{1/2}}$ and ${\vf n}$ is the unit vector along 
particle velocity direction whereas 
$R^{'2}=x^2+(x u)^2=x_{\perp}^2+(({\vf x}{\vf n}) - vx_0)^2/(1-v^2)$. Then the
field strengths are given by
    
\begin{equation}
\label{eq2}
{\vf E}(x)=\frat{e~({\vf x}~u_0-x_0~{\vf u})}
{R^{'3}},~~{\vf H}={\vf v}\times{\vf E}.
\end{equation}
At $v=1$ the potential $A_\mu$ does exist but is singular in the plane
$R^{'}=0$, in particular, moving along $z$-axis it happens to be  that 

$$A_0 = \frat{e}{|z - x_0|},$$
$$A_x = A_y = 0,$$
$$A_z = \frat{e}{|z - x_0|}.$$

The field strengths disappear (equal to zero) in the limit $v \to 1$,
besides the plane $z=x_0$ where the transverse components tend to blow up

$$
E_z = 0,~~E_x = e~u_0~\frat{x}{\rho^3},~~
E_y = e~u_0~\frat{y}{\rho^3},
$$
here with $\rho^2=x^2+y^2$.
Actually, nonexisting finite limit of physical observable quantity, like
the field strengths in this case, signifies simply that a real physical
charge can't be accelerated up to the light velocity because of infinite
amount of energy necessary to do so.

The potential of randomly distributed charges with the center of mass
ultrarelativistically moving in the direction of vector ${\vf n}$ is given
by the following integral over the charge density distribution in the
rest frame of $q({\vf z})$,i.e.

\begin{eqnarray}
\label{eq3}
A_\mu(x)=\int d{\vf z}~ q({\vf z})~ A_\mu(x, {\vf z}),\nonumber\\ [-.2cm]
\\[-.25cm]
A_\mu(x, {\vf z})=\frat{u_\mu}{R^{'}(x,{\vf z})}~~,\nonumber
\end{eqnarray}
where $R^{'2}=-x_0^{2}+({\vf x}-{\vf z}^{'})^2+[-x_0~u_0+
({\vf x}-{\vf z}^{'},{\vf u})]^2$,
and the vector ${\vf z}^{'}$ absorbs the Lorentz contraction of the
system of charges in the factor $u^{-1}_0$ in the motion direction, thus  
${\vf z}^{'}=({\vf z}{\vf n}){\vf n}/u_0+{\vf z}
-({\vf z}{\vf n}){\vf n}.$

In the particular case of a dipole with charges  $e$ and $-e$
being placed at the points ${\vf z}_1=-{\vf d}/2,~{\vf z}_2={\vf d}/2$ 
of the dipole rest frame, the electric field components become 
 
\begin{eqnarray}
\label{eq4}
&&{\vf E}_\bot=e~u_0\left\{\frat{{\vf x}_\bot+{\vf d}_\bot/2}{R_1^{'3}}
-\frat{{\vf x}_\bot-{\vf d}_\bot/2}{R_2^{'3}}
\right\},~~\nonumber\\ [-.2cm]
\\[-.25cm]
&& E_{||}=\frat{e}{R_1^{'3}}\left\{u_0({\vf x}{\vf n})-vu_0x_0+
\frat{({\vf d}{\vf n})}{2}\right\}
-\frat{e}{R_2^{'3}}
\left\{u_0({\vf x}{\vf n})-vu_0x_0-\frat{({\vf d}{\vf n})}{2}\right\}
,~~\nonumber
\end{eqnarray}
with $R_i^{'}=R^{'}(z_i),~~i=1,2$.

Although these fields are singular in the limit $v \to 1$, nevertheless 
one may notice their orthogonal (relative to the motion direction)
components are developing two symmetric peaks opposite-orientated at    
$|{\vf d}_\bot|=0$ . Let us trace now the reaction of fast moving particle  
passing through the field of such a configuration. Due to the Lorentz
contraction this particle is exposed to two very short $\delta$-like
pulses of opposite signs. If the contraction is large enough that the time
interval between two pulses $r_i \sim T$ is less than the classical radius
of particle $a=e^2/m,~d\sim a$ with mass $m$, there is no enough time to
respond both separate pulses and the charge is sensitive to the smeared
(averaged in time) field only. Indeed, the evolution of charge radiation
is controlled by the Newtonian equation with the radiative friction 

$$
\dot w=g(t)+2 e^2 \ddot w/3m,
$$
and its general retarded solution is 
 
$$
w=w_0+\int_{-\infty}^{t} g(\tau) d\tau+\int^{\infty}_{t} g(\tau) 
e^{(t-\tau)/\kappa}d\tau,
$$
if $g(t)$ is the impulse generated by an external field, $\kappa=2e^2/3m$. 
When two $\delta$--like contra-peaks of the amplitude $\sim G$ separated
in time $\sim T$ at the condition $T/\kappa\ll 1$ are present, the
particle velocity $w$ differs from the initial value $w_0$ by the quantity
of $G~T^2/\kappa^2$ order (in the order of $T/\kappa$ the pulses cancel each 
other). Since $G\sim vu_{0}G_0$ and $T\sim T_0/vu_0$ with $G_0$ and $T_0$
being fixed in the rest frame, then in the limit $v \to 1$ it occurs that 
$G~T^2/\kappa^2\to 0.$  Hence, it is well grounded to write down for 
the general solution
 
$$
\label{eq7}
w=w_0+\int_{-\infty}^{t} 
\langle g(\tau)\rangle
 d\tau+\int^{\infty}_{t}
 \langle g(\tau)\rangle 
e^{(t-\tau)/\kappa}d\tau,
$$
where $\langle g(\tau)\rangle$ is an impulse smeared in time.
 
Thus, as the regularized field we should treat one averaged over the
interval $[x_{0i},~x_{0f}]=[({\vf x}{\vf n})/v-({\vf d}{\vf n})/2vu_0, 
~({\vf x}{\vf n})/v+({\vf d}{\vf n})/2vu_0]$, i.e. defined by

\begin{equation}
\label{eq8}
\langle {\vf E} \rangle=\frat{vu_0}{({\vf d}{\vf n})}\int^{x_{0f}}_{x_{0i}}
{\vf E}~ d x_0.
\end{equation}
In particular, if the charge distribution takes such a form that 
$\langle{\vf d}_\bot\rangle\sim 0$ (averaging over charge density), then
we have 

\begin{equation}
\label{eq9}
\langle E_{||}\rangle =\frat{e}{({\vf d}{\vf n})}\left\{
\frat{1}{|{\vf x}_\bot|}-\frat{1}{[({\vf d}{\vf n})^2+{\vf x}_\bot^2]^{1/2}}
\right\},~~
\langle E_\bot\rangle \approx 0.
\end{equation}

It is interesting to notice that our procedure of regularizing by
averaging is similar, in a sense, to the potential regularization as 
done in Ref.\cite{9} dealing with the light-cone variables and then widely
used \cite{5},\cite{6},\cite{10}. If we integrate now the transverse field of 
a singled charge over $dz$ (when moving in $z$-direction) on the infinitesimal
interval $[z_i,~z_f]=[z-vx_0-b/u_0, z-vx_0+ b/u_0]$ ($v \to 1$)
$$
\int_{z_i}^{z_f} E_{\perp}dz =
\frat{e~x_{\perp}}{\rho^2}\cdot \frat{2b}{(b^2 + \rho^2)^{1/2}}.
$$
at large enough value of $b \gg \rho $ then it is valid 
$$
\lim_{\rho/b \to 0} \int_{z_i}^{z_f} E_{\perp}dz \to
\frat{2e~x_{\perp}}{\rho^2},
$$
i.e. the transverse field may be presented in the form of $\delta$ -
function $E_{\perp} = \frat{2e~x_{\perp}}{\rho^2} \delta (z-x_0)$, what is 
equivalent to the equation (8) of Ref. \cite{9}. It was calculated there
starting from the initial potential of Eq.(1), omitting the retarded part
($x_+=(z+x_0)/2 \sim 0$) and performing the singular gauge transformation
of the longitudinal potential components to the transverse ones. Strictly
speaking it's not permissible to drop out the $x_+$-dependence because the 
corresponding terms are of the same order in (1- $v$) as ones used in the 
further calculations of Ref. \cite{9} (the entire transverse field is zero
in this approach as well if we consider the dipole configurations of
longitudinal orientation). For the transverse dipoles
$$
{\vf E}_{\perp}=2e~ \delta (z-x_0) \left\{ \frat{ {\vf x}_{\perp}+
{\vf d}_{\perp}/2}{({\vf x}_{\perp}+{\vf d}_{\perp}/2)^2 }-
\frat{ {\vf x}_{\perp}-{\vf d}_{\perp}/2}{({\vf x}_{\perp}-{\vf d}_{\perp}/2)^2 } \right\}
$$
and we obtain the entire transverse field equal to zero again but after
averaging over its orientation in ($x,y$)-plane for a homogeneous
(like in KR of Ref. \cite{6}) distribution.

\section{ Radiation of fast charged particle in averaged 
field of relativistic dipole }

The radiation energy of a fast moving classical charge in an external
electromagnetic field is determined as \cite{11}
$$
\triangle \varepsilon=\int^{\infty}_{-\infty}dt~
\frat23\frat{e^4_1}{m^2}\frat{({\vf E}+{\vf v_1}\times{\vf H})^2-
({\vf v_1}{\vf E})^2}{1-{\vf v_1}^2},
$$
where $e_1$ and ${\vf v_1}$ are the charge and velocity of radiating
particle, respectively. Here we are interested in the situation when a
particle and an ultrarelativistic dipole oriented along its motion
direction are going towards each other. In accordance with previous
consideration we should substitute the average field instead of instant
values of field strengths in the radiation formula. Such replacement
results in disappearing the singular part of radiation from the transverse
components of field, because $ \langle E_\bot\rangle \approx 0,$
$ \langle H_\bot\rangle \approx 0.$
Then the same factors ($1-v^2_1$) available in the numerator and
denominator are cancelled out and the energy radiated per unit time comes
about to be independent of the particle energy

\begin{eqnarray}
\label{eq11}
\frat{d \Delta \epsilon}{d t} =
  \frat{2e_1^4}{3m^2}~
  \langle E_{||}\rangle^2.
\end{eqnarray}
here we neglect, of course, a particle influence on a dipole in the first
approximation.

For comparison we give the radiation intensity of a fast charged particle
in the field of immobile dipole with its orientation along the direction of 
particle propagation in eikonal approximation:
$$
\triangle \varepsilon = \frat{3~\pi~e_1^4~e^2~d^2 }{64~m^2~\rho^5~v_1
}~(~3 + \frat{5}{1-v_1^2}~)
$$
(see also a similar exercise for the radiation in the field of Coulomb
center \cite{11} ). This radiation is composed by two different
components, i.e. the singular part of transverse components and the 
regular one of longitudinal components. In fact, we can neglect the
longitudinal radiation in the limit $v_1 \to 1$. Our averaging procedure and 
substitution of $ \langle E \rangle$, $ \langle H \rangle$ in radiation
formula instead of $E$ and $H$ kill a singular radiation and we must take 
into consideration small longitudinal radiation 'surviving'. This
substitution arises as a leading term of perturbative expansion in
$T/\kappa$ which we are able to argue basing on the Newtonian equation 
(though it can be done in the framework of relativistic dynamics). Any
other terms should be omitted in the solution of dynamical equation with
radiation force included. Let's trace now the consequences of averaging
procedure.

Integrating Eq.(7) over the impact parameter we have 

\begin{eqnarray}
\label{eq11a}
\frat{d \Delta \epsilon}{d t} = \int \frat{d^2 k_{\perp}}{(2 \pi)^2}
~\frat{2e_1^4}{3m^2}~
E^2_{||}(k_{\perp}),
\end{eqnarray}
where 
\begin{eqnarray}
\label{eq12}
E_{||}(k_\perp)= \int  d{\vf x}_\bot~e^{-i({\vf k}_\bot{\vf x}_\bot)}
~\langle E_{||}\rangle = \nonumber\\ [-.2cm]
\\[-.25cm]
\frat{4 \pi e}{d~k_{\perp}}\Bigg\{1 - \sqrt{\frat{2}{\pi}} (d~k_{\perp})^{1/2}
K_{1/2} (d~k_{\perp})\Bigg\}, \nonumber
\end{eqnarray}
\noindent 
and $K_{1/2}(z)$ is the modified Bessel-function. Then asymptotic
behaviours of the field are the following 

$$
\lim_{k_{\perp} \to 0} E_{||}(k_{\perp}) \to 4 \pi e~, \qquad
\lim_{k_{\perp} \to \infty} E_{||}(k_{\perp}) \to 
\frat{4 \pi e}{d~k_{\perp}}\{1 - e^{-d~k_{\perp}}\} ~.
$$

It results in the finite value of the distribution of radiated photons as
obtained from Eq.(8) divided by the photon energy $k_0$, and being
proportional to the Fourier component of electric field squared, of
course, in the limit $k_{\perp} \to 0$ . It approaches exponentially the
perturbative behaviour $\sim 1/k_{\perp}^2$ at the transverse momenta 
while large enough. Let's emphasize the rate of transition to the
asymptotic regime is regulated by the distance between the dipole charges in 
its rest frame $\sim m^{-1}_{\pi}$ and the radiation energy is
proportional to ($e^6$). 

The $x$ distribution ($x$ is a portion of longitudinal momentum carrying
by an individual parton) takes well-known 'soft' form 
$dk_3/k_0  \approx dx/x$, because 
$\delta(z - x_0) \Rightarrow k_0 =k_3 \gg k_{\perp}^2$. In  order to get
beyond this approximation we need to keep the next order terms of the
($1-v^2$)-expansion. Moreover, in lieu of the coherent summation of the fields
as in Eq.(5) one should average them taking into account the phase shifts
between different points. Calculating with the corresponding phase factor
taken (moving along the $z$-direction) leads to 

$$\langle E_{\perp}e^{ik_3z}\rangle 
= \frat{e}{d}~ 2ik_3 ~e^{ik_3vx_0}~ x_{\perp} 
\frat{ \left(1 + \frat{d^2}{\rho^2}\right)^{1/2} - 1 -\frat{d^2}{2\rho^2}}
{ (\rho^2 + d^2)^{1/2}},$$
$$
\langle E_ze^{ik_3z}\rangle
=\frat{e}{d} ~e^{ik_3vx_0}~\Bigg[\frat{1}{\rho} - 
\frat{1}{[\rho^2+d^2]^{1/2}}\Bigg].
$$
These expressions make it clear that the transverse field (unlike the 
longitudinal one) may be dropped out when one estimates the radiation
behaviour in low transverse momentum region (large impact parameters $\rho$).
Although the transverse field doesn't identically disappear and allows us
to take the limit $v \to 1$, it's justified, nevertheless, because of
the additional suppression $d^{2}/\rho^{2}$ present comparing to the 
longitudinal field. The coherent regime is associated with the constraint
$k_3d < u_0$ , which is apparently valid in the limit of $v \to 1$ for any
$k_3$. We understand, however, these delicate problems require the full
detail investigations in future. 

In fact, the analysis performed may be extended to non-abelian theory.
As shown in Ref. \cite{12} at the small enough coupling constant
$g^2/4\pi\le1$ non-abelian theory is steady similar to the electrodynamics as 
to the problems of both dipole and induced radiation description. The 
influence of non-linear terms, in particular, of 'three gluon vertex'
mainly discussed, is reduced for the dipole configuration to provide
only pure rotation of the vectors of particle colour charges in the
'isotopic' space (for SU(2) non-abelian group) which is certainly
unobservable and doesn't lead to any physical effects. What one should
modify only is to replace the charge product $e_1$ $e_2$ of the interacting 
particles by the scalar product $(\tilde{P_1}$$\tilde{P_2})$ of the
vectors of particle charges in the 'isotopic' space and to take also into
consideration the energy of gluomagnetic field generated by the chromoelectric
charges  $\tilde{P_1}$ and $\tilde{P_2}$ . Thus, the existence of additional
distribution over the relative angles of 'isotopic' vector charges of
quark-antiquark pairs in nuclei foresees major distinction of non-abelian
theory from electrodynamics in such an analysis.

\section{Conclusion }

Using the classical electrodynamics as an example we have demonstrated
above the limit of $v \to 1$ does exist for the physical observable
quantities, what are the average field strengths of dipole configurations
oriented along their propagation direction ('longitudinal' dipole), and
this qualitative picture does not change when non-abelian theory is
considered. In a sense it justifies some intuitive representations about
the structure of classical gluon fields generated by ultrarelativistic
nuclei collision and about nuclear structure functions developed by
McLerran and Venugopalan in \cite{5}. The gluon transverse momentum 
distribution here obtained agrees qualitatively with the distribution presented
in Fig.1 of Ref.\cite{10} though essential difference is that we obtain
the infrared stable and finite result at $k_{\perp}\to 0$ and the
transition to the perturbative regime occurs at the transverse momenta 
$k^2_{\perp} \sim d^{-2} \sim m^2_{\pi}$. We are free to take the minimal
size of hadron as the mean distance between dipole charges in its rest
frame. Unfortunately, we are unable now to point out the well grounded
mechanism to orientate the colour dipoles along their propagation but we
believe (and may argue) the rather plausible hypothesis is that the dipoles 
are aligned along the external accelerating field because such configurations 
in electrodynamics are energetically more favourable. 

Let us summarize now our main results. We suggest the regularization
procedure of classical fields as being generated in ultrarelativistic
heavy ion collisions and relate the field smearing to the Lorentz
contraction of field configurations. If characteristic field configuration
'size' is less than the classical particle radius then this particle is
able to 'feel' an average field only. Operating with the physical
observables we have immediately calculated the average field of fast
moving 'longitudinal' dipole (Eq.\ (6)) in the limit $v \to 1$ and then
the radiation energy of a fast moving charge in this field (Eqs.\ (8), (9)).
The transverse momentum distribution of radiated photons happens to be
finite at small values and exponentially runs into the perturbative regime
$\sim 1/k^2_{\perp}$ at larger $k^2_{\perp}$. Applying proper  
quark-antiquark configurations for the calculation of structure functions
leads to the same result as for the photons distribution and agrees
qualitatively with result obtained earlier in Ref. \cite{10}. 
Unlike the light-cone method every stage of our approach is physically
meaningful because we are dealing with observable values only. Seems, 
including next order terms of expansion in ($1-v^2$) could illuminate the
origin of structure function behaviour at small $x$ what we see as the subject
of our further investigation. 

The work has been initiated by numerous discussions of the 
McLerran-Venugopalan model with V.Goloviznin, M.Gyulassy, Yu.Kovchegov, 
A.Kovner, A.Leonidov, A.Makhlin, L.McLerran, D.Rischke, H.Satz and R.Venugopalan.
The financial support of RFFI (Grants 96-02-16303, 96-02-00088G, 97-02-17491)
and INTAS (Grant 96-678) is greatly acknowledged.

\end{document}